\documentclass{aa}  
\usepackage{graphicx}
\usepackage{txfonts}
\usepackage{soul}
\usepackage[colorlinks=true, citecolor=blue]{hyperref}
\usepackage{natbib,twoopt}
\usepackage{diagbox}
\bibpunct{(}{)}{;}{a}{}{,}             
\usepackage{hyperref}
\hypersetup{
    colorlinks=true,
    linkcolor=blue,
    citecolor = blue,
    filecolor=magenta,
    urlcolor=blue}
\usepackage{booktabs}
\usepackage{lscape}
\usepackage{float}
\usepackage{xcolor}
\usepackage[toc,page]{appendix}
\usepackage{cuted}
\usepackage{caption}

\begin{document}

\authorrunning{Bartolomei et al.}
\titlerunning{The Bulge Cluster Origin (BulCO) survey}

\title{The Bulge Cluster Origin (BulCO) survey with CRIRES at the ESO-VLT: a chemical screening of the Globular Cluster NGC 6553\thanks{Based on observations collected at the Very Large Telescope of the European Southern Observatory at Cerro Paranal (Chile) under Large Program 110.24A4 (PI:Ferraro)}}

\author{A. Bartolomei\inst{1,2}, L. Origlia\inst{2}, C. Fanelli\inst{2}, L. Chiappino\inst{1}\fnmsep\inst{2}, F.R.Ferraro\inst{1}\fnmsep\inst{2},
B. Lanzoni\inst{1}\fnmsep\inst{2}, C. Pallanca\inst{1}\fnmsep\inst{2}, M. Loriga\inst{1}\fnmsep\inst{2}, M. Cadelano\inst{1}\fnmsep\inst{2}, D. Romano\inst{2},  
E. Dalessandro\inst{2}, D. Massari\inst{2}, E. Valenti\inst{3,4}}

   \institute{Dipartimento di Fisica e Astronomia, Università degli Studi di Bologna, Via Gobetti 93/2, I-40129 Bologna, Italy  \email{alessia.bartolomei4@unibo.it}
   \and
   INAF, Osservatorio di Astrofisica e Scienza dello Spazio di Bologna, Via Gobetti 93/3, I-40129 Bologna, Italy
        \and
        European Southern Observatory, Karl-Schwarzschild-Strasse 2, 85748 Garching bei Munchen, Germany
         \and
         Excellence Cluster ORIGINS, Boltzmann-Strasse 2, D-85748 Garching Bei Munchen, Germany}

\abstract
{In this paper we present the chemical screening of the stellar population belonging to the globular cluster NGC~6553 in the Galactic bulge. This study has been conducted in the contest of the {\it Bulge Cluster Origin (BulCO)} survey, an ESO-VLT Large Program currently ongoing with CRIRES in the NIR domain. This survey is performing an unprecedented chemical screening of 17 stellar systems orbiting the Milky Way bulge, with the aim of unveiling their origin and true nature. Here we present and discuss the abundances of 18 elements produced via distinct nucleosynthetic channels for 14 red giant branch stars belonging to NGC 6553. We found a mean [Fe/H] = $-0.20$ $\pm$ 0.01 dex, and about solar-scaled iron-peak elements, confirming that this is one of the most metal-rich globular clusters in the Milky Way.
We also found [X/Fe] enhancement of $\alpha$ and several other light elements.
Furthermore, we assess the presence of multiple populations typical of genuine globular clusters from the significant spreads in Na, N, and C, and an almost vertical Na-O anti-correlation.
Finally, by using classical ([$\alpha$/Fe] vs [Fe/H]) and newly-defined ([V/Fe] and [Zn/Fe] vs [Fe/H]) "chemical DNA tests", we prove its in-situ formation within the Galactic bulge. 
}


\maketitle

\section{Introduction} 
\label{intro}
Galactic globular clusters (GCs) are true touchstones for astrophysics and probably the most studied cosmic objects. With ages of $\sim$ 12 Gyr (e.g., \citealt{Vandenberg13}; \citealt{Valcin20}), they are fossils of the formation epoch of the Milky Way and their study provides crucial insights into the evolutionary history of our galaxy (e.g., \citealt{Searle78}; \citealt{massari19}).
The best studied GCs so far are those orbiting the Milky Way halo, because the prohibitive crowding conditions and the large extinction in the bulge direction have made difficult the exploration of the stellar systems hosted in this portion of the Galaxy (see e.g. \citealt{ortolani97}, \citeyear{ortolani99}, \citeyear{ortolani2001}, \citeyear{ortolani2007}; \citealt{valenti04}, \citeyear{valenti10}; \citealt{cohen21}; \citealt{Souza24}). Still, the Galactic bulge represents the first massive structure to have formed in the Milky Way, and it is the only spheroid in which the kinematic, photometric, and chemical properties of individual stars can be studied. Therefore, the analysis of its stellar clusters provides a unique opportunity to explore the formation and early evolutionary processes of the Galaxy.

Fortunately, the difficulties of observing the Galactic bulge have been minimized in the recent years by the introduction of high-spatial-resolution detectors in the near-infrared (NIR) domain. 
This has given rise to novel research initiatives that use NIR observations to study bulge stellar systems in detail.
For instance, the bulge Cluster APOgee Survey (CAPOS) takes advantage of  the APOGEE-2S spectrograph to obtain precise chemical abundances and kinematic measurements of a substantial number of bulge GCs (\citealt{geisler21}, \citeyear{geisler25}; \citealt{romero21}; \citealt{fernandeztrincado22}; \citealt{barrera25}; \citealt{Frelijj25}; \citealt{gonzalezdiaz23}; \citealt{henao25}; \citealt{uribe25}).
On our side, we are carrying out a project to fully characterize bulge stellar systems by combining high-resolution and multi-wavelength photometric and spectroscopic observations, proper motion (PM) membership selection, and an accurate correction for the effects of differential reddening, combined with a detailed characterization of the extinction law in the direction of each cluster (see, e.g., \citealt{saracino_15,saracino16,saracino19, pallanca19,  pallanca_21, pallanca21b, pallanca23, deras23,loriga25}).
 \begin{figure*} [!ht]
 \sidecaption
    \includegraphics[scale=0.54]{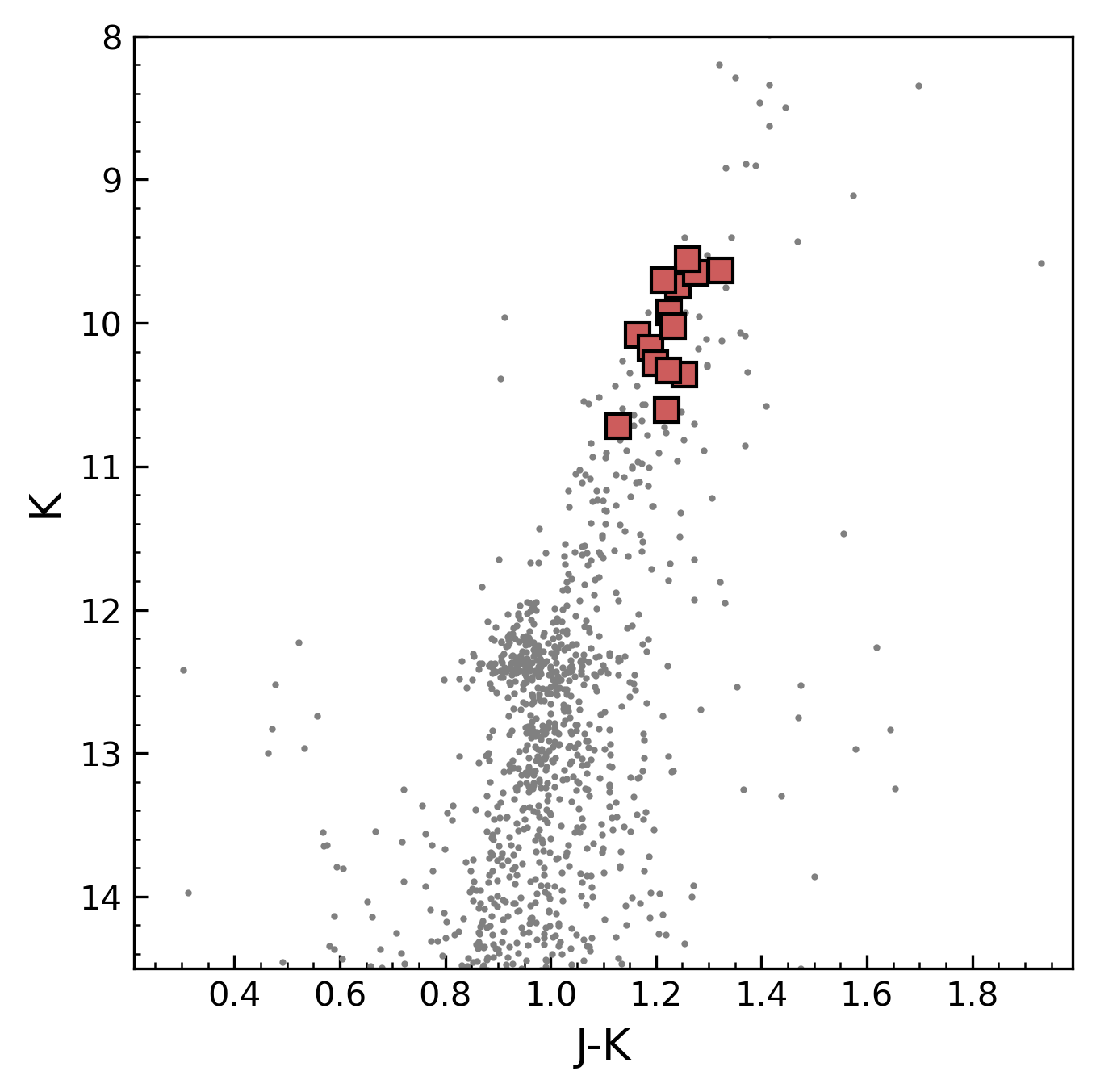} \includegraphics[scale=0.54]{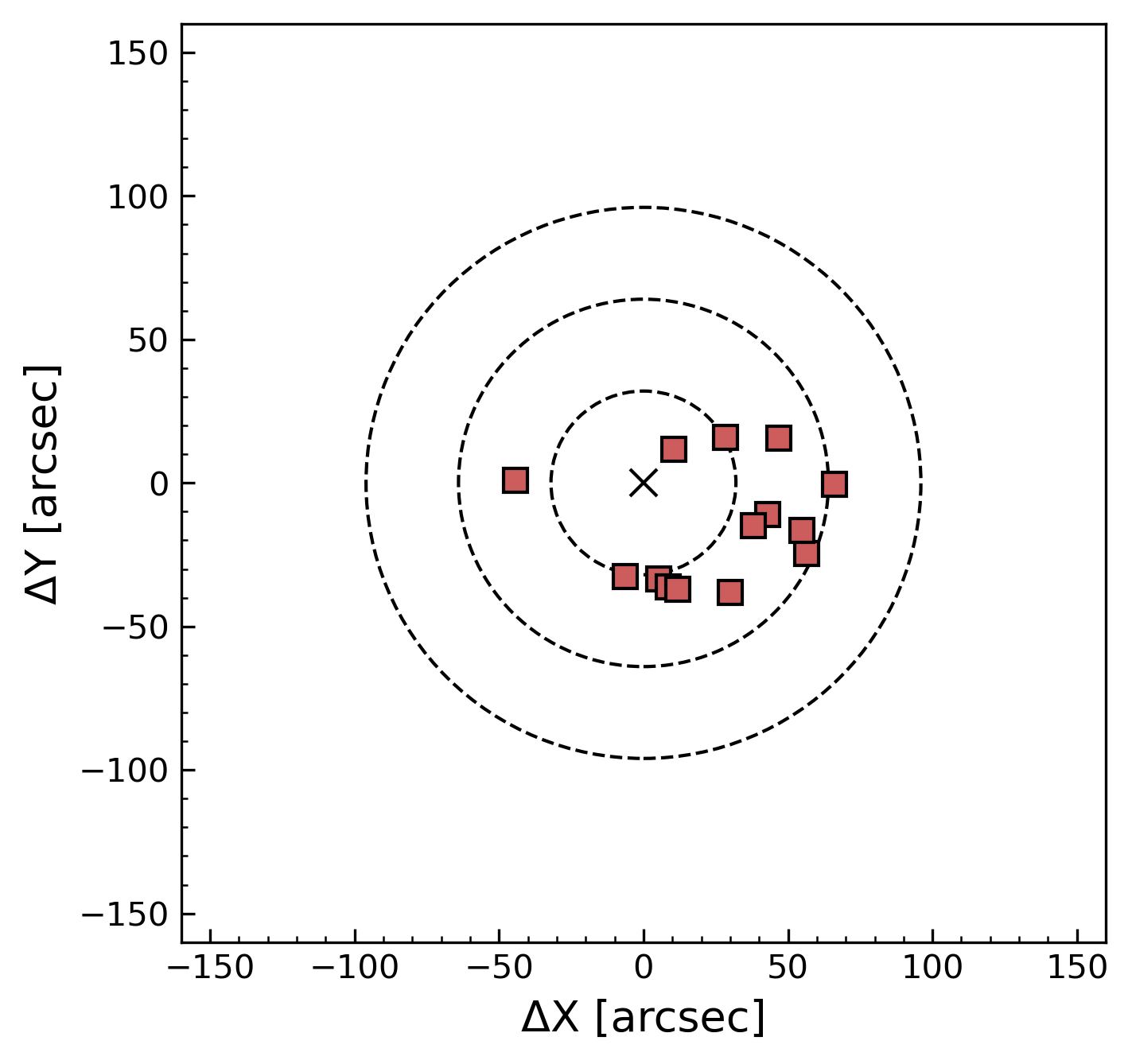}
    \caption{Spectroscopic targets (red squares) observed in  NGC 6553. Left panel: 
    (K, J-K) CMD of the cluster (gray dots) with  highlighted the target stars for which we measured chemical abundances (from \citealt{valenti07}). Right panel: Position of the targets in the plane of the sky with respect to the cluster center (cross). The dashed black circles have radii equal to one, two, and three times the core radius, which is $r_{c}=0.53'$ 
    (\citealt{harris96}).}
    \label{cmd}
\end{figure*}

This exploration has led to the discovery and the characterization of two  
peculiar stellar systems with the appearance of genuine GCs but hosting multi-age and multi-iron sub-populations: Terzan 5 \citep{Ferraro_09, ferraro_16} and Liller 1 (\citealp{ferraro_21}; see also \citealp{pallanca_21}).
Their anomalous properties suggest that they are self-enriched structures (\citealt{romano23}) that experienced complex star formation histories 
(see \citealt{dalessandro_22, crociati+24}). 
Their [$\alpha$/Fe]-[Fe/H] patterns are incompatible with those observed in local dwarf galaxies and in the Milky Way halo and disk, while they are in striking agreement with that of the bulge \citep{Origlia_11, Origlia_13, origlia_19, origlia+25, Massari_14,  crociati_23, Souza23, deimer24,fanelli+24, ferraro+25}. All this suggests that they are not genuine GCs but rather the remnants of massive primordial structures that contributed to form the Galactic bulge \citep{ferraro_21} and that likely are the most efficient factories of gravitational wave sources in spirals (see \citealt{ferraro26b}).

In addition to these fossil remnants, which trace the very first phases of the bulge formation process, and to in-situ formed genuine GCs, a variety of other objects mapping different phenomena are expected to populate the bulge and appear as old stellar systems: genuine GCs that formed in an external galaxy and were brought into the bulge by early and massive accretion events (e.g., \citealt{Massari26}), and possibly nuclear star clusters of cannibalized structures. The signatures of different origins are imprinted in the physical and chemical properties of these stellar systems. Hence, a detailed chemical screening is needed to address the origin of each stellar system. Indeed, this is the  aim of the {\it Bulge Cluster Origin} (BulCO) survey (\citealt{ferraro+25}) which has been specifically designed to perform a high-resolution spectroscopic screening of the stellar population hosted in 17 GC-like stellar systems orbiting the Galactic bulge. It combines the high spectral resolution of CRIRES ($R=50,000$), with the large aperture of the VLT telescope. The survey will provide a homogeneous measure of the chemical abundances of $\sim 20$ different elements for 20-25 stars in each cluster. The abundances of iron, iron-peak, neutron-capture, CNO, and other $\alpha$- and light-elements will be measured and used as chemical DNA tests to distinguish the true nature and origin of each target. In fact, stars formed in environments with different star formation rates (SFRs) are expected to show different chemical patterns (such as in the [$\alpha$/Fe]-[Fe/H] diagram), so distinctive that they can be used as a sort of DNA test to assess the stellar population origin (\citealt{Horta20}, \citealt{Monty24}, \citealt{ceccarelli+24}). The target stars are all located in the very central region of the host stellar system, thus maximizing their membership probability, and span a significant portion of the RGB, thus sampling a large range of  luminosities. These are almost unique features in the chemical investigation of bulge GCs in the NIR domain. In fact, other past and ongoing surveys, performed either with echelle or with multi-object spectrographs, typically at smaller aperture telescopes, are significantly more limited in: (1) the number of observed stars per cluster, which is typically $<10$; (2) the radial location of the targets, which is mostly restricted to the outer cluster regions, where field contamination is much more severe; (3) the spanned luminosity/temperature range, which is typically limited only to the brightest and coolest giants near the RGB-tip, whose spectra are more affected by molecular blending and blanketing; (4) the number of measured chemical elements, which is typically a dozen at most, thus missing crucial diagnostics. All these limitations make it difficult to use the obtained results as solid constraints to the formation and evolutionary processes of  each system, especially in the case of complex scenarios as for Terzan 5 and Liller 1.   

In this paper, we present the results of the chemical screening of the bulge, metal-rich GC NGC 6553. In Sect.~\ref{prev} we present the results of previous chemical studies of this system. In Sect.~\ref{data} we describe the target selection, the assessment of their membership through PMs, and the performed observations.
The estimated stellar parameters and radial velocity measures are reported in Sect.~\ref{rv}. The chemical analysis and the resulting chemical abundances are described in Sect.~\ref{chem}. Finally, in Sect.~\ref{conclu} we discuss the results and draw our conclusions. 

\section{Previous chemical studies of NGC 6553}\label{prev} 
NGC~6553 is one of the most metal-rich GCs in the Galactic bulge, located at 2.2 kpc from the Galactic center ($\alpha_{J2000}~=~18^{h}09^{m}17.6^{s}$, $\delta_{J2000}~=~-25$°$54'31.3''$, l~=~5.26°, b~=~$-$3.03°; \citealt{Harris2010}) and at 6.0 kpc from the Sun. It presents a reddening E(B-V)=0.63 (\citealt{Harris2010}), an age of approximately 11 Gyr (\citealt{ortolani25}), and a mass of ($2.35\pm0.19)\times10^{5}$ $M_{\odot}$ (\citealt{baumgardt_18}; roughly corresponding to an absolute $V-$band magnitude $M_V=-7.7$), which would place NGC6553 in the intermediate-mass range of the Galactic GC distribution.

Several chemical studies have been conducted on NGC~6553 due to its relatively loose and uncrowded structure, as well as its considerably lower reddening compared to the majority of bulge GCs.
\citeauthor{Barbuy92} \citeyearpar{Barbuy92} estimated a metallicity [Fe/H]~$\approx$~$-0.2$ from CCD echelle spectra of a cool giant at a resolution R$\sim20,000$. Subsequently, \citeauthor{Barbuy99} \citeyearpar{Barbuy99} measured [Fe/H] = $-0.55$ and an overabundance of $\alpha$-elements for two member stars.
\citeauthor{Cohen99} \citeyearpar{Cohen99} analyzed 5 horizontal branch (HB) stars using higher resolution HIRES spectra (R $\sim 34,000$) and found a mean value of [Fe/H] = $-0.16$ and an excess of $\alpha$-elements with respect to the solar value. \citeauthor{origlia_02} \citeyearpar{origlia_02} analyzed two cool giants from infrared spectra in the H band at moderate resolution (R $\sim 25,000$) and determined a mean metallicity of [Fe/H] = $-0.3$ and [$\alpha$/Fe] = +0.3.
Later, \citeauthor{melendez03} \citeyearpar{melendez03} studied five giant stars using high-resolution infrared spectra (R$\sim$50,000) in the H band, acquired with the Gemini-South telescope, finding a metallicity of [Fe/H]$ = -0.20$ and an overabundance in oxygen with respect to the solar value from infrared OH lines.
Similar results were found by \citeauthor{Alves-brito06} \citeyearpar{Alves-brito06}, who performed a detailed abundance analysis of four giants using optical high-resolution (R$ \sim 55,000$) echelle spectra obtained with UVES at the ESO VLT-UT2 Kueyen telescope. These are the highest resolution observations in the optical band available for this cluster.
In addition, \citeauthor{johnson_14} \citeyearpar{johnson_14} used high-resolution FLAMES/GIRAFFE spectra (R $\sim 20,000$) to study the behavior of light-, $\alpha$- and Fe-peak element abundances of 12 likely member stars, finding trends nearly identical to those of bulge field stars with comparable metallicity. Similar results regarding the metallicity of NGC 6553 were recently obtained by \citeauthor{schiavon17} \citeyearpar{schiavon17} and \citeauthor{Tang2017} \citeyearpar{Tang2017} through the chemical analysis of red giants using APOGEE's spectroscopy, by \citeauthor{Munoz20} \citeyearpar{Munoz20} through the study of seven member stars using the VLT/UVES spectrograph, and by \citeauthor{montecinos21} \citeyearpar{montecinos21} with the extensive spectral analysis of 20 HB stars. These latest chemical analyzes also reveal the presence of anticorrelations between pairs of light-elements and, consequently, the existence of multiple populations within this cluster.

\begin{figure}
\centering
     \includegraphics[scale=0.75]{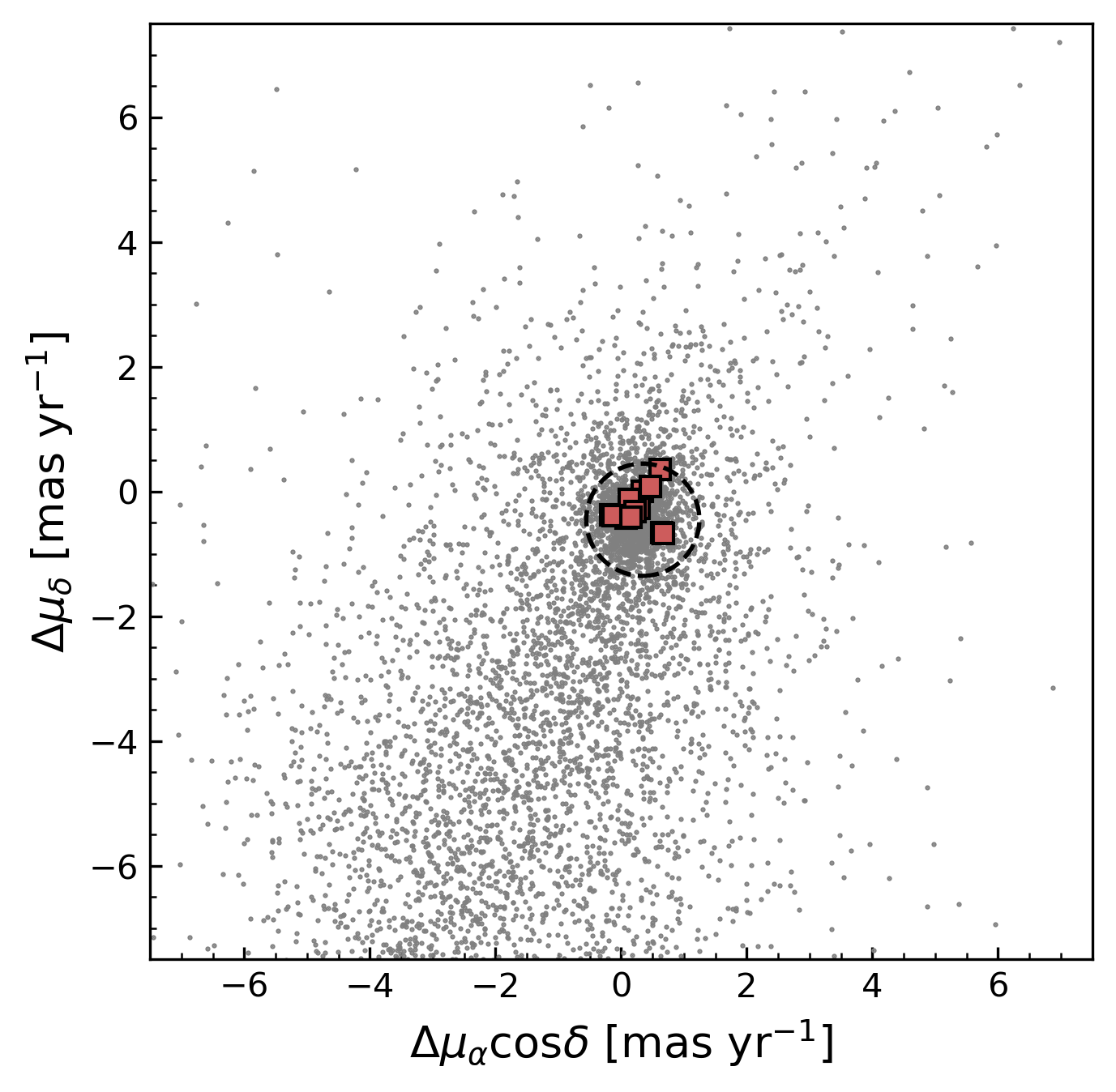}
     \caption{ 
     Vector point diagram obtained from Gaia DR3 PMs, for the stars located within 200" from the center of NGC~6553 (gray dots). The 13 spectroscopic targets with measured PM are reported as red filled squares. The dashed circle has a radius equal to $3 \times \sigma_{PM}$, with $\sigma_{PM} = 0.3$ mas yr$^{-1}$ being the PM dispersion of NGC 6553 member stars. All the spectroscopic targets with measured PM are located within it and have consequently been included in this study.}

     \label{pm}
\end{figure}

\begin{table}
\caption{Coordinates, $J$-band and $K$-band magnitudes, atmospheric parameters and heliocentric radial velocities of the observed stars in NGC~6553.} 
\label{tab1}
\scriptsize
\setlength{\tabcolsep}{4.8pt}
\renewcommand{\arraystretch}{1.9}
    \begin{tabular}{|c|c|c|c|c|c|c|c|}
    \hline\hline
    ID &  RA   &  Dec   & J&  K & $T_{eff}$   & log(g) &RV \\
    \hline
       & [Deg] & [Deg]  & [mag]& [mag]  & K  & dex &km s$^{-1}$ \\
    \hline
        95 &   272.3248194 &  -25.9180200  & 10.980 & 9.739 &   3900 &       1.25 & -0.1$\pm$1.1\\
    \hline
        99 &   272.3109954 &  -25.9084578  & 10.928 & 9.653 &   3850 &       1.25 & -3.1$\pm$0.9\\
    \hline
        108 &   272.3262867 &  -25.9054159  & 10.815 & 9.555 &   3900 &       1.25& 16.3$\pm$0.8 \\
    \hline
        111 &   272.3316673 &  -25.9193165  & 10.916 & 9.702 &   3900 &       1.25 &  13.3$\pm$0.9\\
    \hline
        122 &   272.3390259 &  -25.9154943  & 11.248 & 10.083 &   4000 &       1.25  & -3.6$\pm$1.0\\
    \hline
        130 &   272.3417044 &  -25.9088092  & 11.151 & 9.926 &   3950 &       1.25 &  1.9$\pm$1.0\\
    \hline
        131 &   272.3257339 &  -25.9187631  & 11.360 & 10.171 &   4000 &       1.25 & 6.6$\pm$0.8\\
    \hline
        136 &   272.3352777 &  -25.9117827  & 11.254 & 10.022 &   4000 &       1.25 & -2.3$\pm$0.9\\
    \hline
        160 &   272.3215643 &  -25.9177263  & 11.476 & 10.278 &   4000 &       1.50  & -0.9$\pm$0.7\\
    \hline
        170 &   272.3338623 &  -25.9128648  & 11.615 & 10.361&   4050 &       1.50  & -9.5$\pm$0.8\\
    \hline
        180 &   272.3385474 &  -25.9132711  & 11.554 & 10.330 &   4050 &       1.50 & -1.8$\pm$0.8\\
    \hline
        181 &   272.3312146 &  -25.9043099  & 10.955 & 9.632 &   3800 &       1.00  & -6.6$\pm$1.0\\
    \hline
        204 &   272.3266113 &  -25.9190126  & 11.848 & 10.720&  4100 &       1.50  & 7.4$\pm$0.9\\
    \hline
        249 &   272.3363072 &  -25.9044028  & 11.827 & 10.607 &   4150 &       1.50 & 2.7$\pm$1.0\\    
\hline\hline
\end{tabular}
\vspace{0.15cm}
\end{table}

\section{Observations and data reduction}\label{data}
The BulCO survey takes advantage of the enhanced performance of the CRyogenic high-resolution InfraRed Echelle Spectrograph (CRIRES; \citealt{kaufl+04}; \citealt{dorn+14}; \citealt{dorn+23}) at the ESO Very Large Telescope (VLT).

The targets were selected using high-resolution photometric data and PM-based cluster memberships. Figure~\ref{cmd} shows the positions of the 14 selected stars in the NIR color-magnitude diagram (CMD) and on the plane of the sky with respect to the center of the system. Our targets are luminous giants, at least 1 magnitude brighter than the red clump. All are located in the central region of NGC 6553 within approximately $1'$ from the center, which is equivalent to twice the core radius ($r_{c}=0.53'$; \citealt{harris96}). Table \ref{tab1} lists the coordinates and the J- and K-band magnitudes (from \citealt{valenti07}) of observed targets.

We used \textit{Gaia} DR3 PMs (\citealt{gaia_16}, \citeyear{gaia_23}), which show a distribution peaked at $\mu_{\alpha}cos\delta$~=~0.344~mas~yr$^{-1}$, $\mu_{\delta}$ = $-0.454$ mas yr$^{-1}$, with a dispersion $\sigma_{PM}~\sim~0.3$~mas~yr$^{-1}$ in both the $\alpha$ and $\delta$ components (\citealt{vasiliev_21}).  We assumed that all the stars with PMs within $3 \times \sigma_{PM}$ from the absolute systemic value are likely members of NGC~6553 (see Figure~\ref{pm}).
No Gaia PM measurement is available for one of the targets (namely, 170). Nevertheless, its position in the CMD, its central location and its measured radial velocity (see Section~\ref{rv}) indicate a high probability of cluster membership. For this reason, we decided to include this star in the final sample. 

High-resolution CRIRES spectra of the 14 giant stars of NGC 6553 have been acquired between April 2023 and July 2024 under favorable sky conditions of clearness and seeing.  
All observations were performed with the 0.4$\arcsec$ wide slit, thus providing an overall spectral resolution of $R~\sim~50,000$. 
For each target, observations were made with gratings J1226~(1116~$-$~1356~nm) and H1582 (1484~$-$~1854 nm). These gratings sample a substantial number of unblended spectral lines of iron, iron-peak elements, neutron-capture elements, $\alpha$-elements, and light-elements, as well as a few dozen OH and CO molecular lines. 

Data reduction was performed using the \texttt{CR2RES} version 1.4.4 pipeline\footnote{\href{https://www.eso.org/sci/software/pipelines/cr2res/cr2res-pipe-recipes.html}{https://www.eso.org/sci/software/pipelines/cr2res/cr2res-pipe-recipes.html}}. After the standard dark- and flat-field corrections, each spectrum was sky-subtracted using nod pairs, calibrated in wavelength using arc lamps, and then extracted using the optimum extraction method. This approach minimizes the loss of spectral resolution, maximizes the signal-to-noise ratio, and efficiently identifies local outliers or defects. 
The resulting 1D spectra were then normalized and corrected for telluric features, in order to be properly compared to synthetic spectra. Telluric spectra were extracted from the TAPAS portal\footnote{\href{https://tapas.aeris-data.fr/}{https://tapas.aeris-data.fr/}}. The signal-to-noise ratio per resolution element of the final spectra is always $\ge$40. 

\begin{figure*}
\centering
    \includegraphics[scale=0.70]{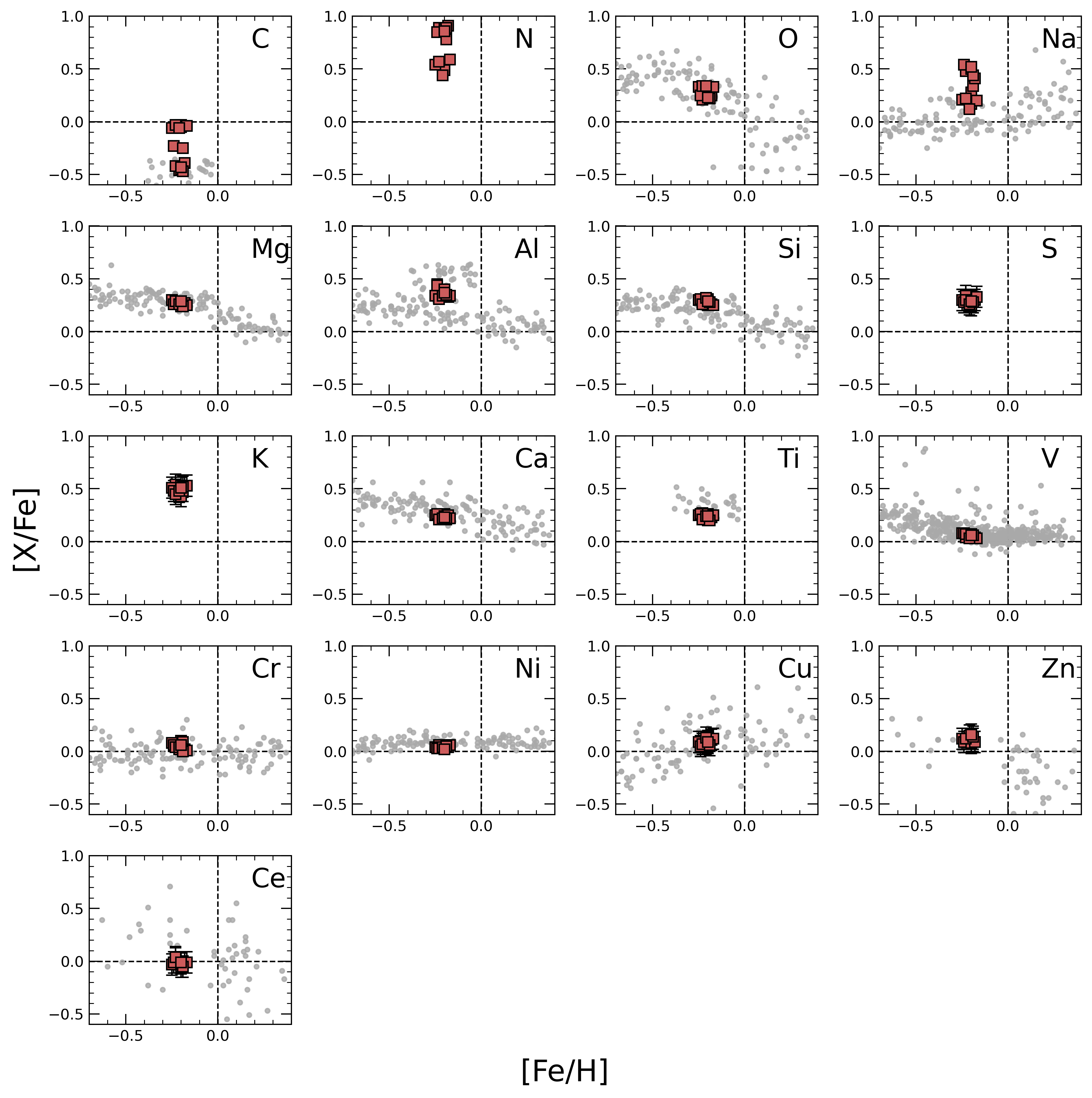}
    \caption{Behavior of iron-peak, neutron-capture, CNO, $\alpha$- and light-element abundances as a function of [Fe/H] for the observed stars of NGC 6553 (red filled squares). The dashed vertical and horizontal lines indicate the corresponding solar [Fe/H] and [X/Fe] values. 
    The small gray circles mark the abundances of bulge field stars as measured by \citet{rich12} for C, O, Mg, Al, Si, Ca, and Ti,  \citet{johnson_14} for O, Na, Mg, Al, Si, Ca, Cr, Ni, and Cu, \citet{Barbuy15} for Zn, and \citet{VanderSwaelmen16} for Ce. The only exception is V, for which the gray circles correspond to disk field stars (from \citealt{Battistini15}).}
    \label{chemestry}
\end{figure*}

\section{Stellar parameters and radial velocities} \label{rv}

An accurate determination of the stellar parameters and radial velocities of the observed stars is a fundamental step of the spectral analysis, since it is preparatory for subsequent measurements of chemical abundances.

A first photometric estimate of surface temperature (T$_{eff}$) and gravity (log $g$) of the spectroscopic targets were obtained through perpendicular projection of each star onto the best-fit isochrone in the CMD shown in Fig.~\ref{cmd}. To this aim, a 11 Gyr old isochrone of appropriate metallicity ([M/H]~$=-$0.1) from the  PARSEC database (\citealt{bressan_12}; \citealt{marigo+17}) has been used.  These initial estimates of the stellar parameters were then spectroscopically refined by fitting the observed molecular OH and CO lines and band-heads simultaneously (see the final values in Table~\ref{tab1}).

Temperatures in the $3800-4150$~K range (with uncertainties of $\pm$100 K), and surface gravities $\log g$ in the 1.00-1.50~dex interval (with uncertainties of  $\pm$0.3 dex) have been estimated. Additionally, a microturbulence velocity of 2 $\pm$ 0.3 km s$^{-1}$ has been adopted for all observed stars. This value  is typical of bulge giant stars with similar temperatures and metallicities (see also \citealt{deimer24}; \citealt{fanelli+24}).

The radial velocities of the 14 RGB stars have been measured by means of cross-correlation between the observed spectra and suitable templates.
The resulting heliocentric values (see Table~\ref{tab1}) range from $-10$ to 16 km s$^{-1}$ with uncertainties around 1 km s$^{-1}$. The error analysis for individual stars was carried out as described by \citeauthor{tonrydavis79} \citeyearpar{tonrydavis79}. For our sample, the mean radial velocity is $1.4\pm1.9$~km~s$^{-1}$ with a 1$\sigma$ dispersion of 7.3~km~s$^{-1}$. This value is in good agreement with other results in the literature reporting the mean velocity of NGC 6553: 4~$\pm$~7.1~km~s$^{-1}$ (\citealt{Cohen99}), 1.6~$\pm$~6~km~s$^{-1}$ (\citealt{melendez03}), $-1.86$~$\pm$~2.01~km~s$^{-1}$ (\citealt{Alves-brito06}), $-2.03$~$\pm$~4.85~km~s$^{-1}$ (\citealt{johnson_14}), 6~$\pm$~8~km~s$^{-1}$ (\citealt{dias15}), $-0.14$~$\pm$~5.47~km~s$^{-1}$ (\citealt{Tang2017}) and $-3.86$~$\pm$~2.12~km~s$^{-1}$ (\citealt{Munoz20}).

\section{Chemical abundances} \label{chem}
In order to measure the chemical abundances, multiple grids of synthetic spectra were generated by adopting the stellar parameters estimated for each star and varying the metallicity from $-1.0$ dex to +0.5 dex, in steps of 0.25 dex. We also considered some enhancement of [$\alpha$/Fe] (+0.3 dex) and [N/Fe] (+0.5 dex) and a corresponding depletion of [C/Fe] ($-0.3$ dex) for a proper computation of the molecular equilibria. We assumed solar-scaled [X/Fe] values for the other elements.
These synthetic spectra have been computed by adopting the list of atomic lines from the \texttt{VALD3} compilation (\citealt{Ryabchikova_15}) and calibrated on Arcturs (see \citealt{Fanelli21}), molecular lines from the website of B.Plez\footnote{\url{https://www.lupm.in2p3.fr/users/plez/}}, MARCS models atmospheres (\citealt{gustafsson_08}), and the radiative transfer code \texttt{TURBOSPECTRUM} (\citealt{alvarez_98}; \citealt{plez_12}). Initially computed at the nominal CRIRES resolution (R $\sim 50,000$), the synthetic spectra were then convoluted with a Gaussian function at R $\sim 29,000$ to account for some additional spectral broadening due to macroturbulence, thus providing an optimal match to the observed line profile.

The chemical abundances of Fe, $\alpha$-elements (Mg, Si, Ca, O, S), light-elements (C, N, Na, Al, K, Ti), iron-peak elements (Ni, Cr, V, Zn, Cu) and neutron-capture elements (Ce) have been determined via spectral synthesis for each selected star.
The final values and their corresponding errors are listed in Table~\ref{results} in the appendix. The formal error is calculated by dividing the standard deviation by the square root of the number of lines used. If only one line is available, a conservative value of 0.10 dex is assumed. However, the global error of each abundance ratio amounts to 0.10-0.15 dex, mainly due to the uncertainties in the atmospheric parameters (especially temperature). The majority of the chemical abundances (Fe, Na, Mg, Al, Si, S, K, Ca, Ti, V, Cr, Ni, Cu, Zn, and Ce) have been determined from atomic lines in the J and H bands. Those of O and N have been derived, respectively, from OH molecular transitions in the H band, and CN molecular transitions in both the J and H bands. C abundance has been obtained from CO molecular transitions and bandheads and from atomic lines in the H-band. Figure~\ref{chemestry} shows the abundance ratios [X/Fe] of all chemical elements measured in the CRIRES spectra as a function of [Fe/H]. An accurate description of these trends is presented in the following sections.

\begin{figure}[]
\centering
    \includegraphics[scale=0.45]{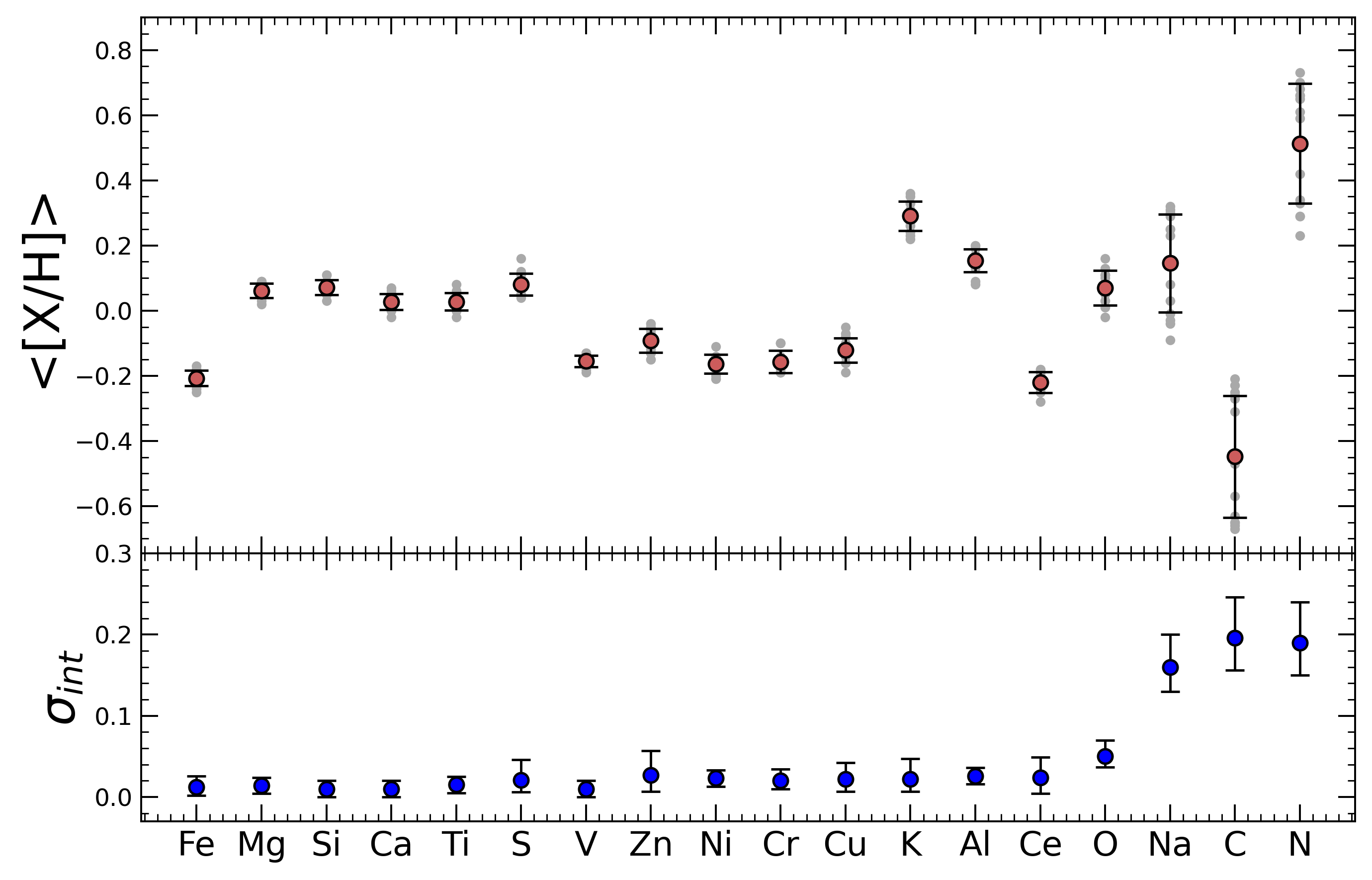}
    \caption{Top panel: Average abundance of each chemical element for the star sample (red filled circles) with their corresponding error. The gray dots refer to the measurements obtained for each individual star. Bottom panel: Intrinsic $\sigma$-value for each chemical element with the corresponding error (blue filled circles).}
    \label{sigma}
\end{figure}

\subsection{Iron and iron-peak elements}
Iron-peak elements are defined as elements with atomic numbers in the range 21~$\leq$~Z~$\leq$~32, from scandium to germanium. These elements are produced through complex nucleosynthesis processes and are divided into two main groups: the lower iron group, with 21~$\leq$~Z~$\leq$~26, and the upper iron group, with 27~$\leq$~Z~$\leq$~32 (\citealt{woosley95}). They can be generated by core-collapse supernovae (SNe), including Type II SNe and hypernovae, as well as Type Ia SNe.

We found an average [Fe/H] = $-0.20$ $\pm$ 0.01 dex, with a 1$\sigma$ scatter of 0.02, which classifies NGC 6553 as one of the most metal-rich GCs of the Galaxy.
The result of the present study is fully consistent within about $\pm$0.1 dex with previous medium-high resolution spectroscopic works (\citealt{Cohen99}; \citealt{melendez03}; \citealt{Alves-brito06}; \citealt{Harris2010}; \citealt{Tang2017}; \citealt{Munoz20}; \citealt{montecinos21}). 

In this study we also measured the chemical abundance of other five iron-peak elements belonging to both groups: V, Cr, Ni, Cu, and Zn. In Figure~\ref{chemestry}, their [X/Fe] abundance ratios are plotted versus [Fe/H], and they are compared with those measured in bulge field stars. In the case of vanadium, the comparison is made with disk field stars due to the unavailability of high-resolution measurements of this element for bulge objects. All these elements show [X/Fe] abundance ratios around solar-scaled values within ±0.10 dex.
V, Cr and Ni consistently trace iron, while Cu and Zn have a more complex nucleosynthesis, and part of their solar abundance is due to weak s-processes in massive stars (\citealt{Bisterzo05}; \citealt{Romano07}).

\subsection{Alpha and other light elements}

In this study we measured chemical abundances for five $\alpha$-elements, namely O, Mg, Si, Ca, S, and for other light elements, namely C, N, Na, Al, K, and Ti.
An average [X/Fe] enhancement  with respect to the solar-scaled values has been measured in the case of $\alpha$-elements, in agreement with other spectroscopic studies of NGC 6553 \citep[see e.g.][]{Cohen99,origlia_02,melendez03,Tang2017,Munoz20} and with the values measured in bulge field stars at a similar metallicity. [X/Fe] enhancements have also been measured in the case of Al, K, and Ti. 
[N/Fe] turns out to be significantly more enhanced with respect to the solar-scaled value, with a bimodal distribution around average values of about +0.55 and +0.85 dex.
The [Na/Fe] ratio also appears to be enhanced with respect to the solar-scaled value, and it clearly shows a significant spread, with values ranging between +0.1 and +0.6 dex.
C is the only light element whose [X/Fe] abundance ratio is normally depleted with respect to the solar-scaled value, with [C/Fe] values ranging between -0.5 and about 0.0 dex. 
We could also measure the $^{12}$C/$^{13}$C isotopic ratio, finding values between 9 and 13. The inferred [C/Fe] depletion and low $^{12}$C/$^{13}$C isotopic ratios are indicative that mixing and extra-mixing processes are occurring during the evolution along the RGB, although, given the relatively small range of stellar temperatures covered by our sample of stars in NGC 6553, we do not find any significant trend with T$_{\rm eff}$.

\subsection{Neutron-capture elements}
Neutron-capture elements are primarily synthesized by two processes: the slow (s) process and the rapid (r) process. In short, the s-process builds heavy elements through slow neutron captures in low-mass AGB stars, as well as fast-rotating, low-metallicity massive stars. In the r-process, instead, the neutron capture rate is much faster than the $\beta$-decay rate of unstable nuclei, which requires high neutron-density environments such as, for instance, magnetorotational-driven SNe, collapsars, or merging neutron stars.
In this paper, we measured the chemical abundance of Ce.
In the solar isotopic mixture, the s-process has been identified as the predominant mechanism for Ce production, with a s/r-process ratio of 85/15 (\citealt{bisterzo14}; \citealt{prantzos20}).  
We measured the Ce abundance using a single ionized line at 15784.786 $\AA$ in the H-band, and the astrophysical $\log gf$ value (transition probability; $g=$ statistical weight, $f=$ oscillator strength) from \citeauthor{Cunha17} \citeyearpar{Cunha17}.
As illustrated in Figure~\ref{chemestry}, the [Ce/Fe] abundance ratio for our NGC 6553 star sample is consistently about solar-scaled.

\section{Discussion and Conclusions}
\label{conclu}

Our CRIRES high-resolution spectroscopic study of 14 RGB stars likely members of NGC 6553 is part of the BulCO survey, an ESO-VLT Large Program specifically aimed at investigating the origin of the most massive and peculiar stellar clusters in the Galactic bulge, and it has included the measurements of precise abundances for 18 chemical elements.

\begin{figure}
\centering
    \includegraphics[scale=0.4]{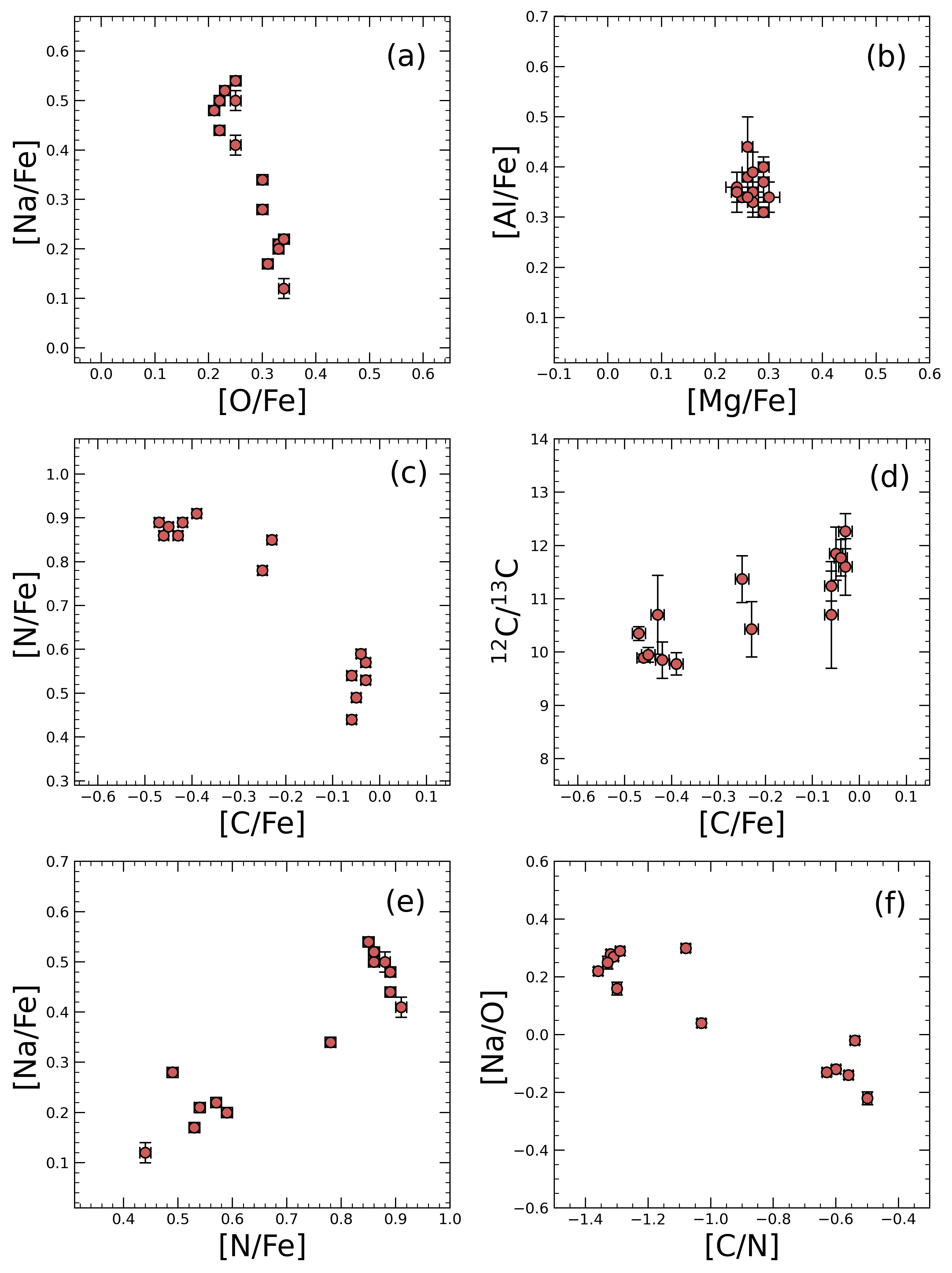}
    \caption{Trends between pairs of light-element abundance ratios for the observed stars of NGC 6553, tracing the presence of multiple stellar populations.} 
    \label{anticorr}
\end{figure}

The inferred iron and iron-peak element abundances consistently trace a high-metallicity of about $-$0.2 dex for this cluster.
An [X/Fe] enhancement of N, O and other alpha and light elements, and a [C/Fe] depletion with respect to the solar-scaled value have also been found.

In order to quantitatively evaluate the presence of elemental abundance spreads within the investigated sample of stars, we computed, for each chemical specie, the average abundance measurement of the 14 stars with its associated error and its corresponding intrinsic spread $\sigma_{int}$ (see the top and bottom panels of Figure \ref{sigma}, respectively).
Most of the elements show a $\sigma_{int}$ spread below 0.03 dex, thus suggesting an homogeneous chemistry within the errors. Notable exceptions are Oxygen ($\sigma_{int}=0.05$ dex) and especially C, N, and Na, which show spreads in the 0.15-0.20 dex range.
Such a behavior is not surprising, since GC stars, unlike galactic field stars, can exhibit significant spreads in light-elements and  specific correlations/anticorrelations (\citealt{carretta09}). The exact origin of these variations is still a matter of debate (e.g., \citealt{Bastian18}; \citealt{Gratton19} for reviews on the subject), but they are commonly believed to have formed during the very early phases of GC formation and evolution (10-100 Myr). The observed trends are suggested to be signatures of the activation of specific nuclear cycles such as the CNO, NeNa, and MgAl modes of hydrogen burning \citep[see e.g.,][]{Arnould99,carretta09}.

The most common chemical signature of the presence of multiple populations, which characterizes nearly all GCs, is the Na-O anticorrelation. The present study reveals a rather steep Na-O anticorrelation in NGC 6553 (Figure~\ref{anticorr}, panel a), with a small spread in O and a rather large spread in Na. 
Another important chemical signature is the Mg-Al anticorrelation, although it is not consistently observed in all GCs, and its presence may depend on the mass and metallicity of the cluster \citep{carretta09b}. We do not find evidence of Mg-Al anticorrelation in our study of NGC 6553 (see Figure~\ref{anticorr}, panel b), consistent with the high metallicity of this cluster.
We also find a quite tight anticorrelation between N and C, with at least one main gap in the distribution between C-rich/N-poor and C-poor/N-rich (Fig.~\ref{anticorr}, panel~c), and  only a modest correlation between $^{12}$C/$^{13}$C and [C/Fe] (Fig.~\ref{anticorr}, panel d).
N-C anticorrelation and some $^{12}$C/$^{13}$C~-~C correlation are usually observed both in field and cluster giants, due to mixing and extra-mixing processes during the evolution along the RGB, and more evolved (i.e. cooler and more luminous) giants typically show more mixing. 
The rather low values of $^{12}$C/$^{13}$C measured in all the sampled stars of NGC 6553 definitely indicate that mixing and extra-mixing are being at work in all of them. 
However, mixing  alone cannot easily explain the presence of the two rather distinct N-poor/C-rich and N-rich/C-poor groups of stars in the N-C plane, given also that all these stars have rather homogeneous stellar parameters and low $^{12}$C/$^{13}$C.
Interestingly, the C-poor/N-rich stars are also the Na-rich ones (see Figure~\ref{anticorr}, panels e and f), suggesting that this group of stars likely formed from gas enriched in N and Na by a previous generation of (N- and Na-poor, C-rich) stars. Hence, all together, the inferred Na-O and N-C distributions seem to provide a solid evidence of the presence of multiple populations in NGC~6553,  thus classifying it as a genuine GC. This result is also consistent with the findings of \citet{Tang2017} and \citet{Munoz20}. 

\begin{figure}
 \centering
    \includegraphics[scale=0.64]{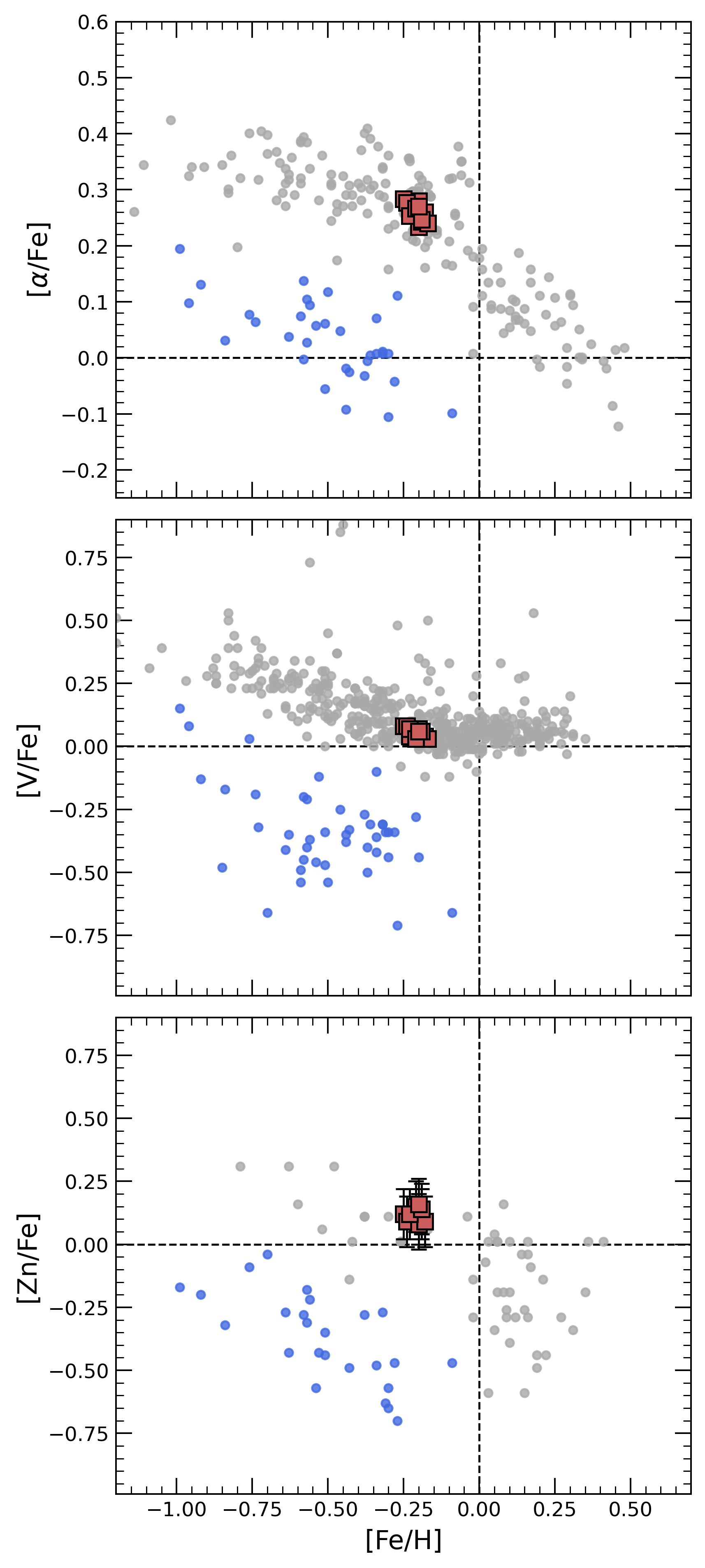}
    \caption{Behavior of [$\alpha$/Fe], [V/Fe] and [Zn/Fe] as a function of [Fe/H] for NGC 6553 (red filled squares). The $\alpha$-element abundance is the average of Mg, Si, and Ca abundances for each star. For comparison, the abundance ratios measured in Milky Way stars (gray circles; \citealt{rich12}; \citealt{johnson_14}; \citealt{Battistini15}; \citealt{Barbuy15}) and in the Magellanic Clouds and Sagittarius  dwarf galaxy (blue circles; \citealt{minelli21a}) are also shown. The abundances of NGC~6553 are consistent with those observed in the Milky Way.}
    \label{alpha}
\end{figure}

From a dynamic perspective, the nature of NGC 6553 remains a subject of debate. In fact, while \citeauthor{massari19}~\citeyear{massari19}, eDR3 edition\footnote{\href{https://www.oas.inaf.it/en/research/m2-en/carma-en/}{https://www.oas.inaf.it/en/research/m2-en/carma-en/}}, and \citeauthor{deleo25}~\citeyear{deleo25} classify it as an in-situ GC, \citeauthor{Callingham22}~\citeyear{Callingham22} classify it as accreted, in particular belonging to the Kraken group. The resolution of this debate can be achieved by incorporating chemical information.

As discussed in \citet[][and references therein]{ferraro+25}, a few specific abundance ratios are able to constrain chemical enrichment timescales and formation scenarios.
In this respect, the [X/Fe] abundance ratios of $\alpha$-elements, Vanadium and Zinc are especially interesting as possible chemical DNA tests \citep[see e.g.,][]{minelli21,mucciarelli21}. 
$\alpha$-elements are primarily produced by Type II SNe (\citealt{woosley95}), with yields depending on the progenitor's mass.
In the metallicity range of NGC~6553, Vanadium and Zinc are mainly produced in high-mass ($\geq20$~M$_{\odot}$) stars, as a result of the explosion of Type II SNe and electron-capture SNe (\citealt{romano10}; \citealt{koba20}; \citealt{palla2021}). Consequently, their abundance ratios are predicted to be larger in high SFR environments, where massive star production is significantly higher compared to low SFR systems. This allows for a clear distinction between in situ-formed and accreted GCs. 
By using Mg, Si, and Ca ($\alpha$-elements that are not expected to show significant spread at high metallicity), we computed an average [$\alpha$/Fe] = +0.26 $\pm$ 0.02 dex enhancement and average [V/Fe] and [Zn/Fe] values consistent with being solar-scaled or barely (if any) enhanced.
As illustrated in Figure \ref{alpha}, our results on NGC 6553 show good agreement with the corresponding abundance patterns of the bulge, and prominent differences with respect to those (significantly lower) measured in the Milky Way satellites.
In addition, considering the cluster’s age and metallicity, NGC 6553 is comparable to the population of in-situ Milky Way GCs, while the stellar systems associated to Sagittarius and the Gaia-Sausage-Enceladus dwarf galaxy exhibit much lower metallicities, on the order of [M/H]$= -1.0$ dex \citep[see][]{Massari26}.

In conclusion, our high resolution chemical study of NGC~6553 indicates that this stellar system is a genuine, metal-rich GC, that formed and evolved in-situ within the Galactic bulge.

\begin{acknowledgements}
This work is part of the project {\it "GENESIS - Searching for the primordial structures of the Universe in the heart of the Galaxy"} (Advanced Grant FIS-2024-02056, PI:Ferraro), funded by the Italian MUR through the Fondo Italiano per la Scienza call.  CF and LO acknowledge the financial support by INAF within the VLT-MOONS and ELT-ANDES projects. ED acknowledges financial support from the INAF Data analysis Research Grant (PI E. Dalessandro) of the “Bando Astrofisica Fondamentale 2024”.
\end{acknowledgements}

\bibliographystyle{aa}
\bibliography{biblio} 

\onecolumn
\begin{appendix}
\section{Chemical abundances}
\begin{table}[H]
\caption{Chemical abundances and corresponding errors of the observed stars in NGC 6553.} 
\label{results}
\scriptsize
\setlength{\tabcolsep}{3.8pt}
\renewcommand{\arraystretch}{1.6}
    \begin{tabular}{|c|c|c|c|c|c|c|c|c|c|c|c|c|c|c|}
    \hline\hline
     \diagbox{[X/H]}{ID} &  95 & 99& 108& 111 & 122 & 130 & 131 & 136 & 160 & 170 & 180 & 181 & 204 & 249 \\
     \hline \hline
     [Fe/H] & -0.21 & -0.20 & -0.23& -0.20& -0.25 & -0.24& -0.19 & -0.20 & -0.18 & -0.23 & -0.17 & -0.21 & -0.19 & -0.20 \\
      & 0.01 (27) & 0.01 (31)& 0.01 (22)& 0.01 (31)& 0.01 (27)& 0.01 (29)& 0.01 (28)& 0.01 (28)& 0.01 (35)& 0.01 (36)& 0.01 (34)& 0.01 (20)& 0.01 (33)& 0.01 (35)\\
      \hline
      [C/H] & -0.67 & -0.23 & -0.65& -0.25& -0.31 & -0.47& -0.44 & -0.65 & -0.57 & -0.26 & -0.21 & -0.27 & -0.66 & -0.63 \\
       & 0.01 (21)& 0.01 (25)& 0.01 (19)& 0.01 (28)& 0.01  (25)& 0.01 (27)& 0.01 (27)& 0.01 (20)& 0.01 (22)& 0.01 (28)& 0.01 (25)& 0.01 (21)& 0.01 (23)& 0.01 (26)\\
       \hline
      [N/H] & 0.65 & 0.33 & 0.66 & 0.29 & 0.29 & 0.61 & 0.59 & 0.68 & 0.73 & 0.34 & 0.42 & 0.23 & 0.70 & 0.66 \\
       & 0.01 (21)& 0.01 (27)& 0.01 (25)& 0.01 (28)& 0.01  (24)& 0.01 (23)& 0.01 (22)& 0.01 (25)& 0.01 (24)& 0.01 (28)& 0.01 (22)& 0.01 (21)& 0.01 (23)& 0.01 (24)\\
       \hline
       [O/H] & 0.01 & 0.11 & -0.02 & 0.10 & 0.08 & 0.01 & 0.11 & 0.05 & 0.07 & 0.11 & 0.16 & 0.13 & 0.03 & 0.03 \\
       & 0.01 (11)& 0.01 (11)& 0.01 (11)& 0.01 (12)& 0.01  (10)& 0.01 (11)& 0.01 (9)& 0.01 (13)& 0.01 (12)& 0.01 (10)& 0.01 (10)& 0.01 (9)& 0.01 (13)& 0.01 (10)\\
       \hline
       [Na/H] & 0.29 & -0.03 & 0.25 & 0.08 & -0.04 & 0.31 & 0.15 & 0.30 & 0.23 & -0.01 & 0.03 & -0.09 & 0.25 & 0.32 \\
       & 0.01 (2)& 0.01 (2)& 0.01 (2)& 0.01 (2)& 0.01  (2)& 0.01 (2)& 0.01 (2)& 0.02 (2)& 0.02 (2)& 0.01 (2)& 0.01 (2)& 0.02 (2)& 0.01 (2)& 0.01 (2)\\
       \hline
       [Mg/H] & 0.06 & 0.04 &  0.03 & 0.07 & 0.05 & 0.02 & 0.08 & 0.09 & 0.09 & 0.06 & 0.08 & 0.05 & 0.05 & 0.09 \\
       & 0.02 (5)& 0.02 (4)& 0.01 (5)& 0.02 (3)& 0.02  (4)& 0.01 (4)& 0.01 (4)& 0.01 (5)& 0.01 (5)& 0.01 (4)& 0.01 (5)& 0.01 (3)& 0.01 (4)& 0.01 (5)\\
       \hline
       [Al/H] & 0.14 & 0.16 & 0.15 & 0.19 & 0.09 & 0.20 & 0.16 & 0.20 & 0.15 & 0.08 & 0.17 & 0.13 & 0.16 & 0.17 \\
       & 0.02 (2)& 0.03 (2)& 0.01 (2)& 0.04 (2)& 0.03  (2)& 0.06 (2)& 0.04 (2)& 0.02 (2)& 0.03 (2)& 0.01 (2)& 0.01 (2)& 0.01 (2)& 0.04 (2)& 0.04 (2)\\
       \hline
       [Si/H] & 0.05 & 0.05 & 0.06 & 0.11 & 0.05 & 0.07 & 0.06 & 0.07 & 0.08 & 0.03 & 0.08 & 0.11 & 0.09 & 0.09 \\
       & 0.01 (8)& 0.01 (8)& 0.01 (5)& 0.01 (5)& 0.01  (7)& 0.01 (6)& 0.01 (8)& 0.01 (5)& 0.01 (7)& 0.02 (4)& 0.01 (4)& 0.01 (4)& 0.01 (6)& 0.01 (8)\\
       \hline
       [S/H] & 0.08 & 0.05 & 0.11 & 0.05 & 0.05 & 0.04 & 0.09 & 0.08 & 0.12 & 0.07 & 0.16 & 0.05 & 0.09 & 0.09 \\
       & 0.10 (1)& 0.10 (1)& 0.10 (1)& 0.10 (1)& 0.10  (1)& 0.10 (1)& 0.10 (1)& 0.10 (1)& 0.10 (1)& 0.10 (1)& 0.10 (1)& 0.10 (1)& 0.10 (1)& 0.10 (1)\\
       \hline
       [K/H] & 0.26 & 0.23 & 0.31 & 0.33 & 0.26 & 0.24 & 0.29 & 0.31 & 0.35 & 0.22 & 0.36 & 0.27 & 0.33 & 0.31 \\
       & 0.10 (1)& 0.10 (1)& 0.10 (1)& 0.10 (1)& 0.10  (1)& 0.10 (1)& 0.10 (1)& 0.10 (1)& 0.10 (1)& 0.10 (1)& 0.10 (1)& 0.10 (1)& 0.10 (1)& 0.10 (1)\\
       \hline
       [Ca/H] & 0.04 & 0.01 &  0.02 & 0.06 & 0.00 & 0.02 & 0.02 & 0.04 & 0.07 & -0.02 & 0.05 & 0.01 & 0.03 & 0.03 \\
       & 0.02 (7)& 0.01 (9)& 0.01 (6)& 0.01 (9)& 0.01  (7)& 0.01 (9)& 0.01 (7)& 0.01 (11)& 0.01 (7)& 0.01 (7)& 0.01 (7)& 0.01 (8)& 0.01 (8)& 0.01 (8)\\
       \hline
       [Ti/H] & 0.01 & 0.00 &  0.04 & 0.04 & 0.00 & 0.02 & 0.03 & 0.06 & 0.05 & -0.02 & 0.08 & 0.03 & 0.01 & 0.04 \\
       & 0.01 (8)& 0.01 (10)& 0.01 (7)& 0.01 (8)& 0.01  (9)& 0.01 (7)& 0.01 (10)& 0.01 (8)& 0.01 (8)& 0.01 (9)& 0.01 (6)& 0.01 (5)& 0.01 (9)& 0.01 (8)\\
       \hline
       [V/H] & -0.16 & -0.16 &  -0.19 & -0.15 & -0.17 & -0.16 & -0.16 & -0.13 & -0.14 & -0.15 & -0.14 & -0.18 & -0.13 & -0.15 \\
       & 0.01 (3)& 0.01 (4)& 0.02 (4)& 0.01 (3)& 0.01  (4)& 0.01 (5)& 0.01 (4)& 0.02 (4)& 0.01 (5)& 0.01 (4)& 0.01 (3)& 0.02 (4)& 0.02 (3)& 0.01 (4)\\
       \hline
       [Cr/H] & -0.19 & -0.10 &  -0.19 & -0.10 & -0.17 & -0.18 & -0.10 & -0.16 & -0.16 & -0.19 & -0.16 & -0.17 & -0.19 & -0.14 \\
       & 0.01 (6)& 0.01 (6)& 0.01 (5)& 0.01 (6)& 0.01  (4)& 0.01 (7)& 0.01 (5)& 0.01 (7)& 0.01 (7)& 0.01 (5)& 0.01 (4)& 0.01 (6)& 0.01 (7)& 0.01 (7)\\
       \hline
       [Ni/H] & -0.15 & -0.14 &  -0.17 & -0.16 & -0.21 & -0.21 & -0.15 & -0.17 & -0.14 & -0.20 & -0.11 & -0.17 & -0.14 & -0.18 \\
       & 0.02 (7)& 0.01 (6)& 0.02 (5)& 0.01 (6)& 0.03  (5)& 0.01 (5)& 0.01 (4)& 0.01 (7)& 0.01 (7)& 0.01 (5)& 0.01 (5)& 0.02 (4)& 0.01 (6)& 0.02 (7)\\
       \hline
       [Cu/H] & -0.14 & -0.12 & -0.14 & -0.12 & -0.16 & -0.19 & -0.13 & -0.10 & -0.07 & -0.16 & -0.05 & -0.08 & -0.13 & -0.11 \\
       & 0.10 (1)& 0.10 (1)& 0.10 (1)& 0.10 (1)& 0.10  (1)& 0.10 (1)& 0.10 (1)& 0.10 (1)& 0.10 (1)& 0.10 (1)& 0.10 (1)& 0.10 (1)& 0.10 (1)& 0.10 (1)\\
       \hline
       [Zn/H] & -0.06 & - & - & -0.10 & -0.13 & -0.15 & -0.07 & -0.12 & -0.09 & -0.11 & - & - & -0.05 & -0.04 \\
       & 0.10 (1)& -& -& 0.10 (1)& 0.10  (1)& 0.10 (1)& 0.10 (1)& 0.10 (1)& 0.10 (1)& 0.10 (1)& - & -& 0.10 (1)& 0.10 (1)\\
       \hline
       [Ce/H] & -0.22 & -0.25 & -0.20 & - & -0.28 & -0.25 & -0.24 & -0.22 & -0.19 & -0.19 & -0.18 & - & -0.24 & -0.21 \\
       & 0.10 (1)& 0.10 (1)& 0.10 (1)& -& 0.10  (1)& 0.10 (1)& 0.10 (1)& 0.10 (1)& 0.10 (1)& 0.10 (1)& 0.10 (1) & -& 0.10 (1)& 0.10 (1)\\ 
       \hline
       $^{12}$C/$^{13}$C & 9.89 & 11.60 & 9.85 & 11.85 & 11.24 & 10.43 & 11.37 & 9.95 & 9.78 & 12.27 & 11.77 & 10.70 & 10.35 & 10.70 \\
       & 0.07 (2)& 0.53 (3)& 0.34 (2)& 0.50 (2)& 0.28 (3)& 0.52 (2)& 0.44 (3)& 0.14 (2)& 0.21 (2)& 0.33 (2)& 0.34 (2) & 1.00 (1)& 0.13 (3)& 0.74 (2)\\ 
    \hline
     \hline
\end{tabular}
\vspace{0.20cm}\\
Notes: The adopted solar reference abundances are from \citealt{magg_22}.
The quoted errors are the standard deviations divided by the square root of the number of lines used (which is reported in brackets), with the exception of a conservative assumption of 0.1 dex when only one line was measurable (in the case of the following species: S, K, Cu, Zn and Ce). The global error of each abundance ratio amounts to 0.10-0.15 dex, due to the uncertainties in the atmospheric parameters.
\end{table}
\twocolumn
\end{appendix}

\end{document}